\documentclass[aps,pre,superscriptaddress,amsmath,amssymb,showpacs]{revtex4}
\usepackage{amsmath}
\usepackage{graphicx}
\usepackage{amssymb}
\usepackage{subfigure}
\usepackage{epstopdf}
\usepackage{dcolumn}
\usepackage{bm}

\makeatletter

\begin{document}

\title{Measurement of the 3-D Born-Oppenheimer Potential of a Proton in a Hydrogen Bonded System
using Deep Inelastic Neutron Scattering: The Superprotonic Conductor Rb$_3$H(SO$_4$)$_2$}

\author{D. Homouz}
\affiliation{Physics Department,
  University of Houston,
  Houston, TX 77204}
\author{G. Reiter}
\affiliation{Physics Department,
  University of Houston,
  Houston, TX 77204}
\author{J. Eckert}
\affiliation{LANSCE, Los Alamos National Laboratory, and University of California, Santa Barbara}
\author{J. Mayers}
\affiliation{ISIS, Rutherford Appelton Laboratory}
\author{R. Blinc}
\affiliation{Stefan Josef Institute, Lubljana, Slovenia}

\date{\today}
\begin{abstract}
\noindent We report the first direct measurement of the proton 3-D
Born-Oppenheimer (BO) potential in any material. The proton potential surfaces
in the hydrogen bonded superprotonic conductor Rb$_3$H(SO$_4$)$_2$ are extracted 
from the momentum distribution measured using Deep Inelastic Neutron Scattering(DINS).  The potential  has a single minimum along the bond direction, which accounts for the absence of the  antiferroelectric   transition  seen in the deuterated material, and  a saddle point off the bond direction for tunneling into the next well with a barrier height of 350 meV. 
 The measured potential is in qualitative agreement with phenomenological double Morse
 potentials that have been used to describe hydrogen bonds in other systems.

\end{abstract}

\maketitle

The shape of the Born-Oppenheimer potential of the proton in  hydrogen bonded systems plays an essential role in a variety of phase transitions, in proton conductivity in biologically and technically important materials,  and in the mechanisms of a wide variety of chemically and biologically interesting reactions.
As such, it has been the subject of extensive experimental, theoretical and phenomenological investigations over many years. It can sometimes be inferred from neutron and  light scattering  measurements of the excitations in the system, but these are rarely sufficient to provide an unambiguous characterization of the potential, particularly if the potential energy surface is highly anharmonic.  Ab-initio calculations of the  potential are subject to  approximations of uncertain validity, and are usually assumed to be accurate when they agree amongst themselves.  We present here the first direct, model independent,  measurement of the  of the 3-D proton Born-Oppenheimer potential in any system, and in particular, in a hydrogen bonded system of technological interest,  Rb$_3$H(SO$_4)_2$. \cite{Haile}  These measurements can be used to investigate the accuracy of ab-initio calculations and provide a data base for their improvement, to interpret neutron and light scattering measurements, and as input to simulations of the dynamics of the system containing the  bond.  In this case, the material is a superprotonic conductor, and the form of the potential is essential information in simulating the transport of the protons. The measurement agrees qualitatively with earlier phenomenological attempts  to characterize the O-H-O bond   using empirical Morse potentials. \cite{Matsushita82} A previous determination of the potential by  means  of inelastic neutron scattering and Raman scattering in the system we have measured\cite{Fillaux} is qualitatively wrong, indicating the difficulty of determining the potential from spectroscopy.  The present measurements, made with   neutron Compton scattering, have an rms error of about 20mev over most of the domain in which the potential can be measured. The technology exists to reduce that to 2mev. 
  
    We  extract the information needed to determine the potential from a  measurement of the proton 
momentum distribution by a procedure suggested by Reiter and Silver.\cite{RS}  In order to do this, the proton must be at a center of inversion symmetry, the interaction with other protons must be negligible,  and the  temperature must be lower than any excitations of the proton vibrations. All of these conditions can be satisfied in  RB$_3$H(SO$_4$)$_2$, which is of interest both as a superprotonic conductor at high temperatures and because of the large isotope effect it exhibits at low temperatures. The deuterated material has an anti-ferroelectric phase transition at 82K, while the protonated version has no anti-ferroelectric transition at all. \cite{Gesi80} An earlier experiment on KDP\cite{RMP} was able to extract the potential along the bond, but it was necessary to assume that the potential was separable because of the insufficient quality of the data. We make no assumptions here about the a priori form of the potential. 

  The geometry of this H-bonded system consists of pairs of sulphate tetrahedra hydrogen bonded together, well separated from other pairs.  The distance between nearest neighbor protons is  $5.07\AA$,  leaving the protons on different bonds effectively isolated from each other. The $A2/a$ space group places the average proton position at a center of inversion. This   allows us to assume that the ground state momentum wavefunction is real.  We will argue later that this is not just a statistical average, and that the hydrogen bond is itself symmetric.  The potential can be expected to change with temperature, and indeed, we observe that it does, but we will only be concerned here with the low temperature region. 

Several experimental studies suggested the presence of  a double well  proton potential along the hydrogen bond
\cite{Fillaux,Dolinsek,Mikac} at same time that other experimental studies as well as
theoretical calculations supported single well potential models \cite{Tachikawa, Noda92, Matsushita82}. 
In Fillaux et al's work\cite{Fillaux}, which investigated several related compounds, the peaks in INS spectra were assigned to a slightly asymmetric potential with a double well shape along the bond direction. We find that the potential is unambiguously a single well along the bond. 
There is a barrier to escape of the proton from the well into neighboring wells off the bond axis, with a barrier height of 
approximately 350 meV, in qualitative agreement  with NMR experiments on the deuterated material. \cite{Dolinsek}. 
Fillaux et al's potential is far too broad to be able to describe our momentum distribution measurements, pointing up the difficulty of inferring potentials from a vibrational spectrum, particularly when the correct assignments are difficult to make.

The momentum distribution is measured with Deep Inelastic Neutron Scattering(DINS), also called Neutron Compton Scattering.  DINS is inelastic neutron scattering in the limit of high momentum
transfer $\vec{q}(30-100\AA^{-1})$. In this limit the impulse
approximation\cite{RS} can be used to interpret the scattering process
which means that neutrons scatter from individua protons in the same
manner that freely moving particles scatter from each other.  Thus the
scattering function depends only on the proton momentum distribution
$n(\vec{p})$.
In the impulse approximation limit, the neutron scattering function is related to the momentum distribution
$n(\vec{p})$ by the relation

\begin{align}
\label{eq:sf}
S(\vec{q},\omega)=\frac{M}{q}\int n(\vec{p})\delta(y-\hat{q}.\vec{p})d\vec{p}=
\frac{M}{q}J(\hat{q},y)
\end{align}

where $y=\omega-q^2/2M$. S($\vec{q},\omega$) is the  Radon transform of
n($\vec{p}$).  The relation 
is invertible by a series expansion method  \cite{RS, RMN} that represents n($\vec p$) in terms of three parameters $\sigma_i$, giving the Gaussian widths in three directions, and a set of anharmonic parameters, a$_{n,l,m}$, giving the deviations from an anisotropic Gaussian distribution.\cite{COM} If the a$_{n,l,m}$ were zero, the potential would be harmonic.

For temperatures that are well below the lowest possible excitation
energy of the proton, $n(\vec{p})$ is determined almost entirely by the
ground state of the proton, as will be the case with our measurements
. The shape of $n(\vec{p})$  is determined by the Born-Oppenheimer potential that localizes the proton.  Its second moment gives the
kinetic energy of the proton. There are in principle corrections to the momentum distribution due to the motion of the oxygen ions, but these are insignificant\cite{Warner83} in harmonic systems as a consequence of the large ratio of the masses of the proton and oxygen. We have verified by numerical calculations that they are insignificant also for the anharmonic potentials using Path Integral Molecular Dynamics (PIMD). \cite{tobepub}

Using a single particle potential to interpret DINS  measurements allows us to relate 
$n(\vec{p})$ directly to the proton ground state wave function  via,

\begin{align}
\label{eq:psi}
n(\vec{p})=\frac{1}{(2\pi\hbar)^3}\left|\int\psi(\vec{r})\exp(i\vec{p}.\vec{r})d\vec{r}\right|^2
\end{align}

 We can take  the momentum wavefunction to be  real if the bond is symmetric, This allows us to calculate the spatial wavefunction as well as the BO potential.  This is given from the one-particle Schroedinger equation by,

\begin{align}
\label{eq:pot}
E-V(\vec{r})=\frac{\int\exp(i\vec{p}.{r})\frac{p^2}{2m}\sqrt{n(\vec{p})}d\vec{p}}{\int\exp(i\vec{p}.\vec{r})\sqrt{n(\vec{p})}d\vec{p}}
\end{align}

For asymmetric bonds, a perturbative approach can be used to extract the symmetric part of the  potential 
surface. In this approach, the anharmonic terms in the momentum distribution are 
related to the anharmonic terms in the Taylor expansion of the potential \cite{RS}.

\begin{figure}[p]
\centering 

	\includegraphics[width=5in]{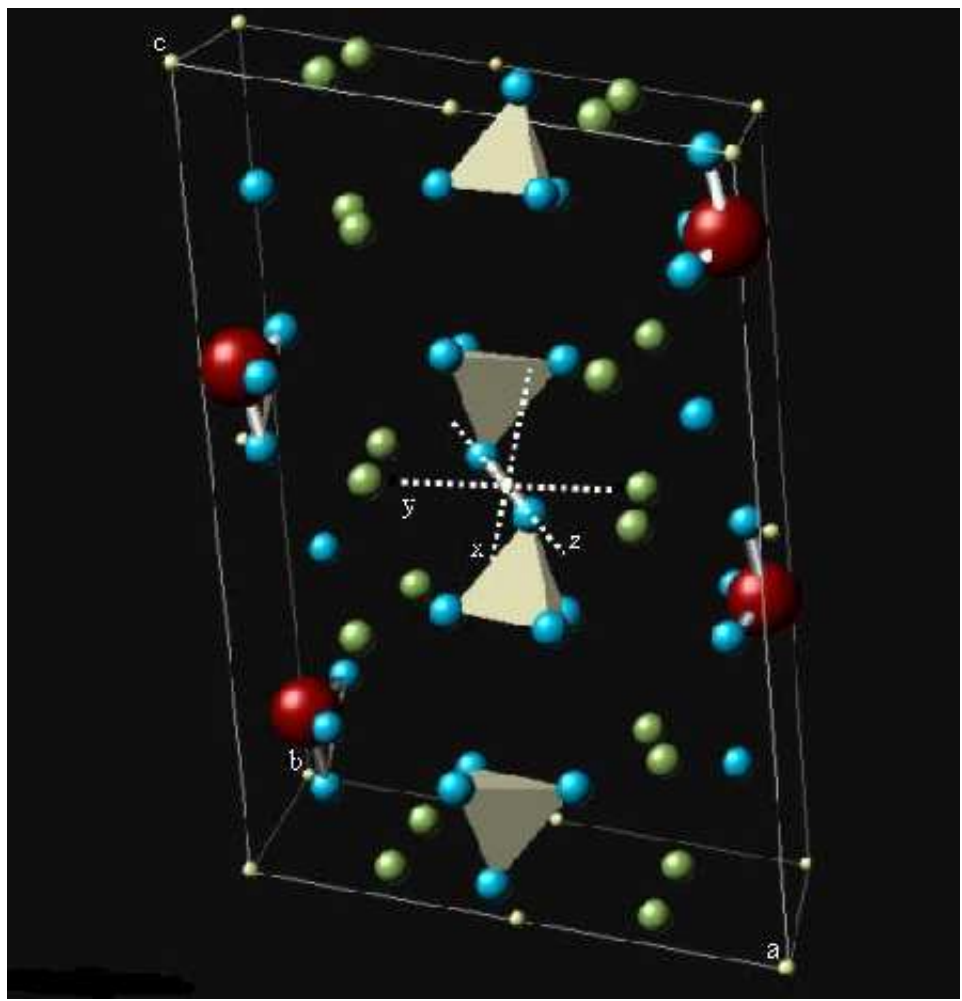}
	\label{fig:3Dstructure}

\caption{(a) The $A2/a$ crystal structure for $Rb_3H(SO_4)_2$ and the orientation of our coordinate system. Oxygen(blue), hydrogen(small white), rubidium(olive), sulphur(red). The gold pyramids are symbols for the SO$_4$ groups.  
The $z-axis$ represents  the H-bond direction, the $y-axis$ is perpendicular to the bond in the ab plane, which is the plane  with significant high temperature proton conductivity. }
\label{fig:structure}
\end{figure}

The DINS experiments were performed on the VESUVIO instrument at ISIS.\cite{Andreani05}
Measurements were done on single crystal samples of Rb$_3$H(SO$_4$)$_2$ at $10K$ and $70K$.  The crystal structure for Rb$_3$H(SO$_4$)$_2$ is shown in Fig.~\ref{fig:structure}, along with the coordinate system that we will use to describe the potential. The z axis is chosen along the bond, the y axis along the line joining the two Rb atoms in the $ab$ plane. Two perpendicular planes of data are collected for the crystal sample. One plane is the $ab$ plane where $b$ is the unique axis of the monoclinic crystal.
 The other plane is the $ac^*$, where the $c^*$ axis is perpendicular to the $ab$ plane.The $ab$ plane is the plane that contains the Hydrogen bond network. The bonds are tilted slightly with respect to it.  The unit cell for Rb$_3$H(SO$_4$)$_2$ has two H bonds at 60$^o$ from each other as shown in Fig.~\ref{fig:structure}. 
 The momentum distribution of each bond is represented as in Refs.(6,11), and the contribution of both bonds added to fit the data.  The measurements were done with 30 detectors in a plane, distributed with scattering angles from 32$^o$ to 68$^o$

\begin{figure}[p]
\centering 
\subfigure[]{
	\includegraphics[width=.45\textwidth,height=.45\textwidth]{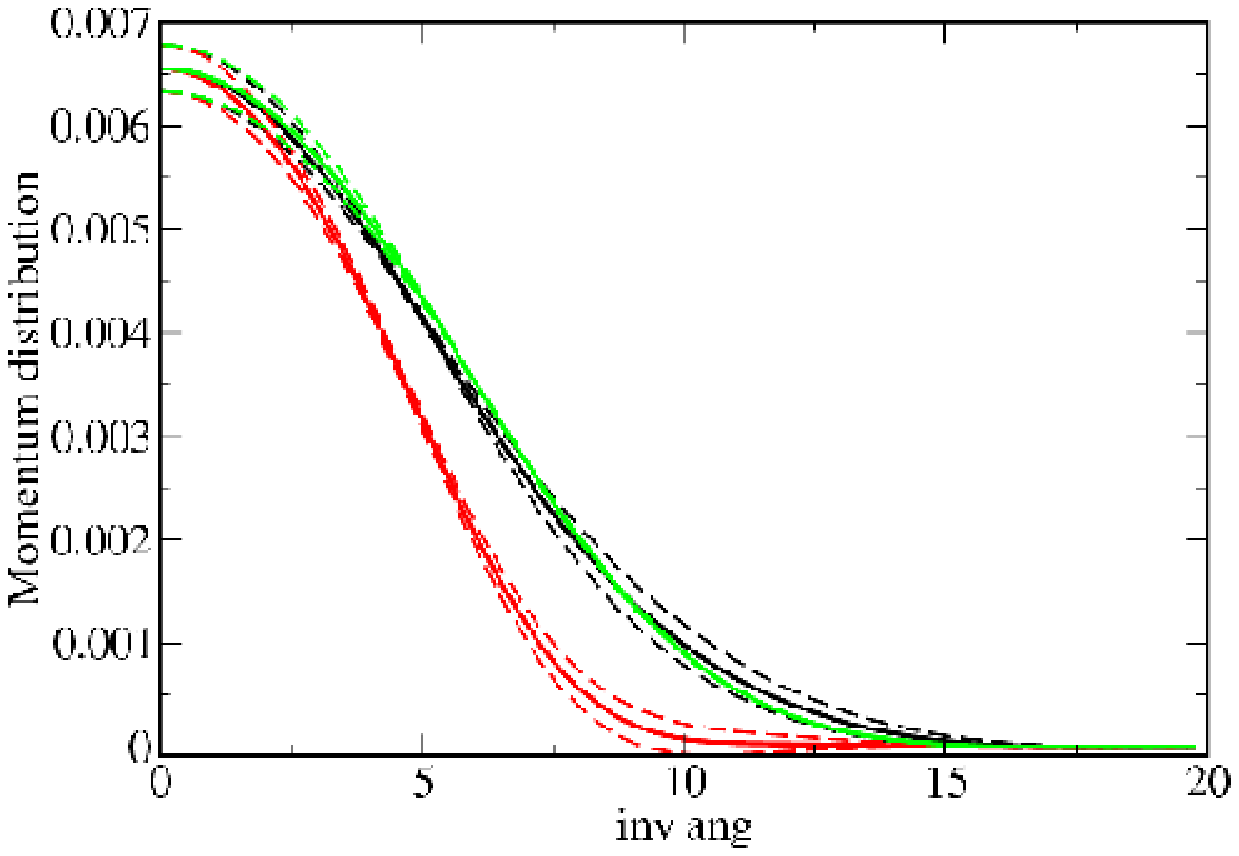}
	\label{fig:mdxyz}
}
\subfigure[]{
	\includegraphics[scale=.6]{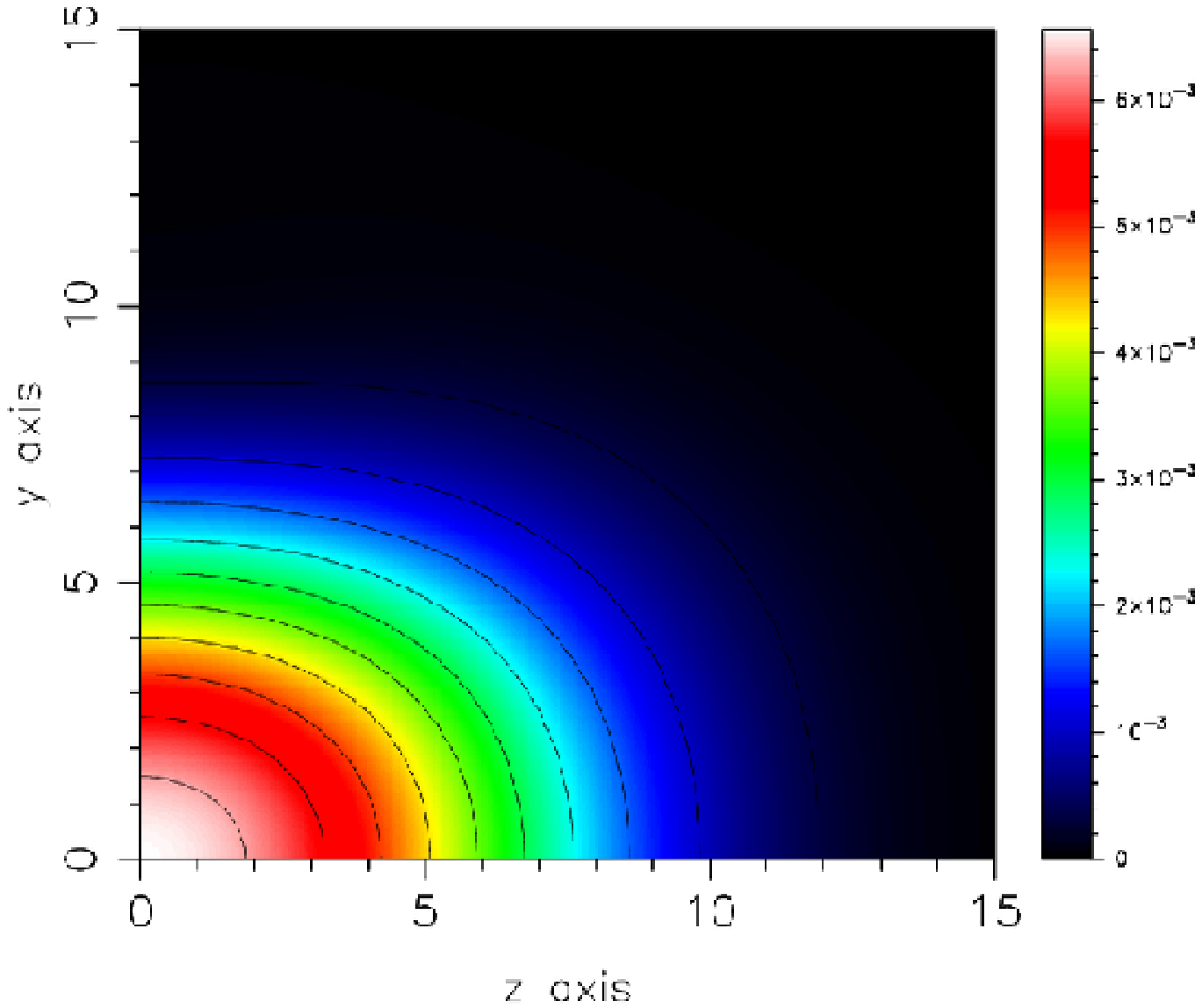}
	\label{fig:mdyz}
}
\caption{Proton momentum distribution for $Rb_3H(SO_4)_2$ at $10K$: 
 (a) along the $x-axis$ (Black), $y-axis$ (Red) and $z-axis$ (Green), the rms errors are shown in dashed lines,
 (b) in the $yz$ plane.}
\label{fig:momdis}
\end{figure}

\begin{figure}[p]
\subfigure[]{
	\includegraphics[width=.45\textwidth,height=.45\textwidth]{figure3a.eps}
	\label{fig:vxyz}
}
\subfigure[]{
	\includegraphics[scale=.6]{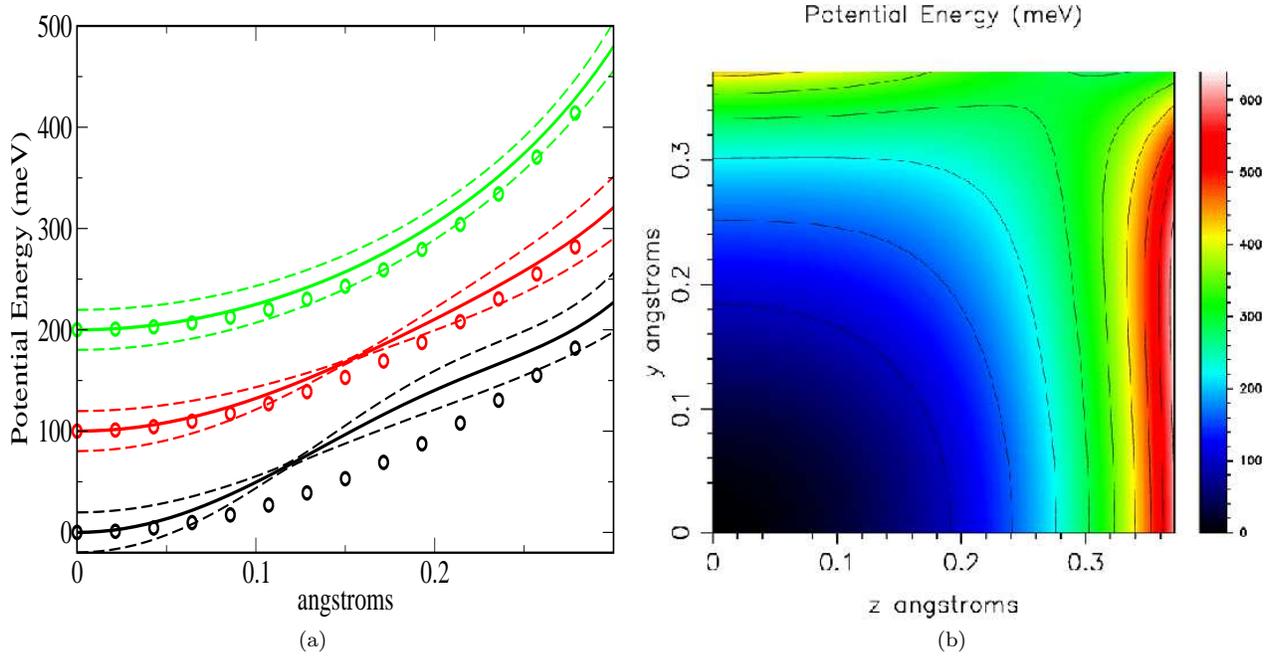}
	\label{fig:vyz}
}
\caption{BO potential for protons in Rb$_3$H(SO$_4)_2$ at $10K$: 
(a) along the $x-axis$ (Black), $y-axis$ (Red) and $z-axis$ (Green), the curves are
shifted by 100 meV along the vertical axis for clarity. The errors are shown in dashed 
lines,The fit
 to a double Morse potential with parameters shown in the first column of  Table III are shown in circles.   (b) The potential energy surface in the $yz$ plane.}
\label{fig:BOpot}
\end{figure}

The fitting parameters for the momentum distribution at the two temperatures are given in tables~\ref{tab:hc} and~\ref{tab:ac}.
Many of these parameters have a confidence level of 3$\sigma$ or higher. 
There is no  phase transition in the protonated Rb$_3$H(SO$_4$)$_2$ for temperatures below 70K. Yet the values of 
these parameters are accurate enough to reveal  some structural changes between 10K and 70K, due simply to the expansion of the crystal.
The proton momentum distribution along the three axes is shown in Fig.~\ref{fig:mdxyz}. Its projection in the $yz$ plane is
 shown in Fig.~\ref{fig:mdyz}. These figures show no sign of proton coherence over two sites, which would show up as an oscillation in the momentum distribution.\cite{RMP} This can be also seen from 
the BO potential surface plot in the same plane  Fig.~\ref{fig:BOpot}. Along the bond direction, $z-axis$, the potential at both temperatures has a single well with a flat bottom. This is shown more clearly in Fig.~\ref{fig:vxyz} where we show the potential in all three coordinate directions, along with the errors\cite{RMN}in the measurement. In general, the measurement can only be done out to distances for which the proton has some significant probability of penetrating. For our present statistics, this is about 2.5 standard deviations of the spatial wavefunction in the direction of interest. 

\begin{figure}[p]
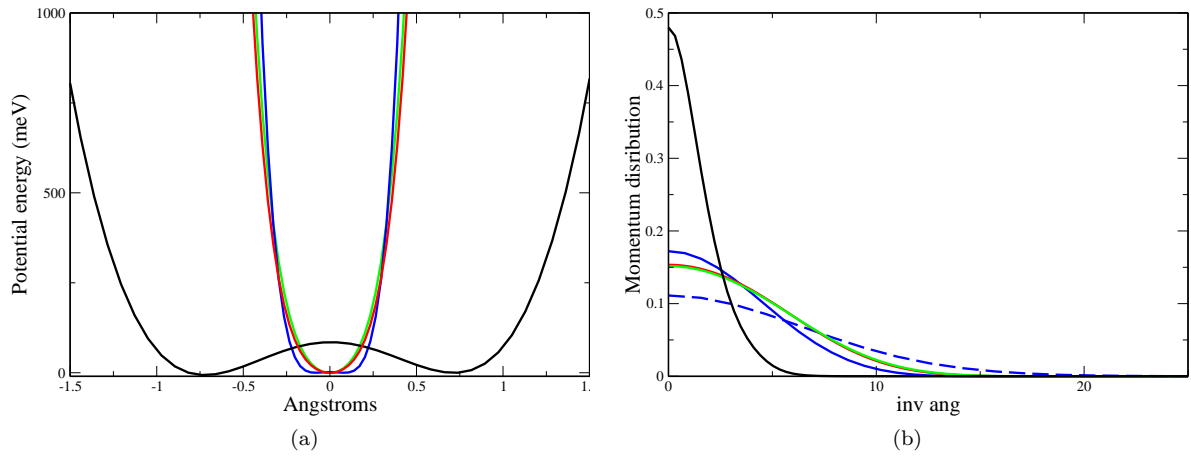

\subfigure[]{
	\includegraphics[scale=.32]{figure4a.eps}
	\label{fig:CompBO}
}
\subfigure[]{
	\includegraphics[scale=.32]{figure4b.eps}
	\label{fig:CompMD}
}
\caption{Comparing NCS measurements with the double-well potential model of Fillaux and a Morse Potential Model.
 Fillaux's model is shown in black. The NCS measurements at 10K and 70K are shown in red and green respectively.
 Matsushita and Matsubara's  model is shown in blue.
  (b) The momentum distributions associated with the potentials in (a). The dashed line is the fit to a single Morse potential}
\label{fig:Comp}
\end{figure}

Although there is no double well along the bond, it is clear that there is an off axis barrier to motion out of the well into the next well.  The barrier height is about 350meV, approximately half the value that was inferred from NMR measurements on the deuterated material.\cite{Dolinsek} at temperatures above room temperature. 																																																																																																																																								In Fig.~\ref{fig:CompBO} we compare this potential with the double-well model suggested by Fillaux {\it et al}\cite{Fillaux}  to interpret the INS measurements for the vibrational spectrum of Rb$_3$H(SO$_4$)$_2$ at $20K$. 
This model is supposed to produce energy levels that can be assigned to the peaks in the INS spectrum. 
Calculating the momentum distribution for Fillaux et al's model and comparing it with our measurement along the bond direction, Fig.~\ref{fig:CompMD}, 
shows that the inferred potential is in fact far from anything that could reproduce our measured momentum distribution, as is also obvious by direct comparison of the measured and inferred potentials.
Ignoring the motion of the heavy ions, and calculating the excitation energies from the measured potential, we find that the first three excitations are at 125 mev, 147 mev, and 155 mev above the ground state at  229 mev, with the zero of energy the zero of the potential. There are only two strong peaks identifiable in the INS spectrum, at  142 mev and 192 mev.  Fillaux et al assigned these to the transverse modes, and identified a band at around 50 mev as the result of the 0-2 transition along the bond interacting with the the SO$_4$ groups, and a mode at  about 5mev as the tunnel splitting. We think in fact that the mode at 192 mev is actually a composite of the (coincidentally) nearly degenerate vibration along the bond and the  transverse vibration in the x direction, which we see at 155 mev and 147 mev respectively.  That these should be close in frequency can be seen directly  from fig 2a and 3a.  The lack of accuracy of our  predicted frequencies is perhaps due to the size of our error bars, which become larger in the region of the excited state energies, where the ground state wavefunction is small.  Since the excited states probe the potential at distances greater than the ground state does, a fitting procedure that combines known spectra with the potential measurements has  the possibility of  extending the range and accuracy of the potential measurements, but we have not  attempted that here.

This single well potential of the  short H bond in the protonated Rb$_3$H(SO$_4$)$_2$ is the reason behind the absence of the low T antiferroelectric transition that exists in the deuterated systems.  It length is $2.4\AA$ by our measurements and 2.483 at room temperature. \cite{Fortier85}
The deuterated H bond, on the other hand,  $2.518\AA$ at 25K \cite{Ichikawa92}, is significantly  longer than the protonated one, and seems to be, in contrast to the situation in protonated Rb$_3$H(SO$_4$)$_2$ , of the double minimum type. 
 Rb$_3$H(SO$_4$)$_2$ contracts upon cooling which reduces the H bond length.
 This suggests that the double minima potential disappears upon cooling in the protonated  system well before the temperature for the onset of the AF transition in the deuterated material. This  is consistent with ab-initio calculations of the structure that take into  account the lowering of the energy by delocalization of the proton.\cite{Tachikawa}, and experimental results on the related material K$_3$H(SO$_4)_2$.\cite{Noda92}

Double Morse potentials (DMP)have been used to describe O-H-O bonds phenomenologically in a variety of physical systems.
\cite{Holzapfel72, Lawrence81, Matsushita82, Mackowiak87, Borgis92, Mashiyama04}  It  is of interest to see how close  they 
actually are to the measured potential. We show in Fig.~\ref{fig:vxyz} the best fitting DMP to our data in the yz plane. The fit includes our fit of the O-O separation.
It can be seen that the fit is semi-quantitatively correct.  We also tried fitting our momentum distribution measurements with a single Morse potential.  This would be appropriate if the symmetric crystal structure were only an average, and individual bonds were actually off center, covalently bonded to one or the other of the oxygens.  That fit doesn't work. When we compare the measured momentum distribution with that of a
single Morse potential of several known models, we see that these models give Gaussian distributions with widths that are significantly bigger than our harmonic parameters along the bond direction. For instance, the models by Holzapfel, 
Matsushita et al, Mashiyama, and Borgis et al have momentum distributions of widths $8.2\AA^{-1}$, $6.6\AA^{-1}$, $7.5\AA^{-1}$ and
$7.15\AA^{-1}$ respectively compared to $4.6\AA^{-1}$ for the harmonic parameter along the bond direction in our measurements. 
The fact that a symmetric DMP fits the measured potential very well and a single Morse potential localizes the proton far too much 
is some indication  that the bond is symmetric. The best fit parameters are collected in Table ~\ref{tab:dmp}  and 
compared with those of Holzapfel\cite{Holzapfel72} and Matsushita and Matsubara \cite{Matsushita82}. The parameters were chosen by Matsushita and Matsubara
to reproduce, in a semi-empirical way, the structural properties of many $OH-O$ complexes. The parameters in both  
works represent the average behavior of wide range of H-bonds.  They don't necessarily represent an isolated H-bond, as in our measurements. A recent study  gives very different parameters for a specific system, KDP\cite{Mashiyama04}, table ~\ref{tab:dmp}, which suggests that there is not an accepted set of parameters. However,  although the parameters appear quite different between our measurement and the parameters of Matsushita and Matsubara, for instance,  the potentials, over the region in which the proton has significant probability of being found, are very similar, up to an additive constant. We show in Fig 4a the comparison of the potential and momentum distribution obtained by  Matsushita and Matsubara when the O-O separation  is set to 2.4$\AA$ with no other changes. 

 Changing the O-O separation in the DMP potential  allows for a transition from single to a 
double well potential as the O-O separation increases beyond a critical value.  The measured critical value for a similar system, K$_3$H(SO$_4$)$_2$ is 
$2.47\AA$ \cite{Noda92} compared to $2.51\AA$ in our DMP fit. Comparing our measurements with Ab-initio Calculations made for isolated Hydrogen bonds in Aquonium Perchlorate \cite{Hud}, we see that a bond length of about $2.4\AA$ is very well into the single well regime. The Path Integral Car-Parinello calculations made by Benoit et al.\cite{Benoit05} give a critical length of about  $2.36\AA$. According to this calculation a bond of a length of $2.4\AA$ has a very low barrier, several meVs. Such barriers wouldn't be noticeable in our measurements within the experimental error. This is true as well of the small barriers and asymmetry  observed in the NMR  measurements of Mikacs et al\cite{Mikac00} 
The single well nature of the bond is therefore consistent with all other available evidence. 

We have shown here the first measurements of a 3-D Born Oppenheimer potential in any system. The measurements can be done on any symmetric hydrogen bond for which the interaction between the protons in different bonds is negligible. The errors in our measurements  are due primarily to counting time limitations on VESUVIO, the one existing machine capable of doing NCS experiments. Its count rate could be improved by a factor of 10 by the addition of more detectors, and another factor of ten could be achieved by the construction of a similar machine on SNS or another newer source. We conclude, therefore, that routine  measurements of Born-Oppenheimer potentials in hydrogen bonded systems to within $\pm 2meV$  are feasible with existing technology. 
\centerline{\bf Acknowledgements:}
The work of G. Reiter and D. Homouz was supported by  DOE Grant DE-FG02-03ER46078. The authors would like to thank Phil Platzman for his comments and a careful reading of the manuscript, and Sossina Haile for useful conversations .

\begin{table}[p]
\caption{The harmonic scale factors for
the momentum distribution measured at 10K and 70K.\label{tab:hc}}
\begin{small}
\begin{tabular}{crrrrcrrrrc}
\hline
\hline
\multicolumn{11}{c}{Harmonic Coefficients}\\
\hline
\hline
    &&&&& 10K &&&&& 70K\\
\hline
i &&&&& $\sigma_i$ ($^o$A$^{-1}$)&&&&& $\sigma_i$ ($^o$A$^{-1}$)\\
z &&&&& 4.60 &&&&& 4.70 \\
x &&&&& 4.35 &&&&& 4.36 \\
y &&&&& 3.73 &&&&& 3.65 \\
\hline
\end{tabular}
\end{small}
\end{table}

\begin{table}[p]
\caption{The anharmonic fitting coefficients and their variances for
the momentum distribution measured at 10K and 70K.\label{tab:ac}}
\begin{small}
\begin{tabular}{crrcrrcrrcrrc}
\hline
\hline
\multicolumn{13}{c}{Anharmonic Coefficients}\\
\hline
\hline
    &&&& 10K &&&&&& 70K &&\\
\hline
$n l m$ &&& $a_{nlm}$ &&& $\delta a_{nlm}$ &&& $a_{nlm}$ &&& $\delta a_{nlm}$\\
4  0 0 &&& -2.05E-01 &&& 2.0E-02 &&&  -1.91E-01 &&& 2.0E-02\\
4  4 2 &&& -8.76E-02 &&& 4.0E-02 &&&  -1.02E-01 &&& 4.0E-02\\
4  4 4 &&& -8.09E-02 &&& 1.0E-02 &&&  -1.26E-01 &&& 1.0E-02\\
6  6 6 &&&  0.0           &&&            &&&  -2.52E-02 &&& 1.0E-02\\
10 6 6 &&&  1.49E-03 &&& 9.0E-04 &&&   2.38E-03 &&& 1.01E-03\\
10 8 8 &&& -3.92E-03 &&& 3.0E-03 &&&  -9.48E-03 &&& 2.0E-03\\
\hline
\end{tabular}
\end{small}
\end{table}

\begin{table}[p]
\caption{Comparison of the measured BO potential, fit with a Double Morse potential(DMP)
of the form $U_{dm}(r)=U(r-z_0)+U(r+z_0)$, where the single morse potentials $U(r)=D\{\exp[-2a(r-r_0)]-2\exp[-a(r-r_0)]\}$, with several other DMP models.[Matsushita and Matsubara(MM), Holzapfel(HZ), Mashiyama(MA)]\label{tab:dmp}}
\begin{small}
\begin{tabular}{lllllllllllllll}
\hline
\hline
    &&&&& This work &&&&&   MM   &&    HZ   &&   MA    \\
\hline
$z_0(^oA)$ &&&&& 1.2$\pm$0.07&&&&& && && \\
$D(meV)$ &&&&&  2920$\pm$205 &&&&& 2150 && 5337&& 2200\\
$a(^oA^{-1})$ &&&&& 1.96$\pm$0.01 &&&&& 2.89 && 2.8$\pm$.2&& 3.8\\
$r_0(^oA)$ &&&&& 0.90$\pm$0.06 &&&&& 0.95 && 0.956 && 1.00\\
\hline
\end{tabular}
\end{small}
\end{table}
\bibliography{bib}

\end{document}